\documentclass[
 aps,
 pra,
 amsmath,amssymb,
 superscriptaddress,
 reprint,
]{revtex4-1}

\usepackage[english]{babel}
\usepackage[utf8]{inputenc}
\usepackage[T1]{fontenc}
\usepackage{amsmath}
\usepackage{amssymb}
\usepackage{graphicx}
\usepackage{bm}
\usepackage{physics}
\usepackage[colorlinks]{hyperref}

\renewcommand{\vec}[1]{\bm{#1}}

\begin{document}

\title{Self-focusing of multiple interacting Laguerre-Gauss beams in Kerr media}

\author{Lucas Sá}
\altaffiliation[Present address: ]
{CeFEMA, Instituto Superior T\'ecnico, Universidade de Lisboa, Av.\ Rovisco Pais, 1049-001 Lisboa, Portugal}
\affiliation{GoLP/Instituto de Plasmas e Fusão Nuclear, Instituto Superior Técnico, Universidade de Lisboa, 1049-001 Lisboa, Portugal}

\author{Jorge Vieira}
\email{jorge.vieira@tecnico.ulisboa.pt}
\affiliation{GoLP/Instituto de Plasmas e Fusão Nuclear, Instituto Superior Técnico, Universidade de Lisboa, 1049-001 Lisboa, Portugal}

\date{\today}

\begin{abstract}
Using a variational approach, we obtain the self-focusing critical power for a single and for any number of interacting Laguerre-Gauss beams propagating in a Kerr nonlinear optical medium. As is known, the critical power for freely propagating higher-order modes is always greater than that of the fundamental Gaussian mode. Here, we generalize that result for an arbitrary incoherent superposition of Laguerre-Gauss beams, adding interactions between them. This leads to a vast and rich spectrum of self-focusing phenomena, which is absent in the single-beam case. Specifically, we find that interactions between different modes may increase or decrease the required critical power relative to the sum of individual powers. In particular, high-orbital angular momentum modes can be focused with less power in the presence of low-orbital angular momentum beams than when propagating alone. The decrease in required critical power can be made arbitrarily large by choosing the appropriate combinations of modes. Additionally, in the presence of interactions, an equilibrium configuration of stationary spot size for all modes in a superposition may not even exist, a fundamental difference from the single-beam case in which a critical power for self-focusing always exists.
\end{abstract}

\maketitle

\section{Introduction}
Lasers carrying orbital angular momentum (OAM)~\cite{allen1992}  have attracted much attention recently due to a vast and interesting set of possible applications, including in plasma-based acceleration~\cite{vieira2014,vieira2016naturecomm}, optical tweezers~\cite{padgett2011,padgett2014}, quantum computation~\cite{mair2001,molina2007}, super-resolution microscopy~\cite{vieira2016prl}, optical communications~\cite{wang2012,bozinovic2013,zhou2018}, imaging~\cite{jack2009}, and astrophysics~\cite{tamburini2011}. Since many applications of OAM depend on the propagation of the beams in nonlinear media it is important to establish the self-interaction of OAM beams as well as the interactions between different OAM modes in those media.

One way to treat the laser-medium interaction is to consider the envelope evolution of the vector potential of the laser beam, which is described by some nonlinear partial differential equation, for instance, the nonlinear Schr{\"o}dinger equation (NLSE) for the paraxial propagation of an ultra-intense short-pulse laser. The optical medium is described by a set of nonlinearities, which can be instantaneous (local along the beam), notably quadratic (Kerr type), or noninstantaneous, breaking translational invariance along the beam. An example of a medium with both types of nonlinearity is a relativistic underdense plasma, where an instantaneous quadratic nonlinearity arises from the relativistic transverse quiver motion of the electrons (relativistic mass correction) and a noninstantaneous one results from the coupling of the laser to the plasma waves. Neglecting noninstantaneous nonlinearities effectively reduces the problem to $(2+1)$-dimensional propagation (meaning that the dynamics along the beam decouple from the dynamics in a plane transverse to it; also, the coordinate along the beam is then essentially equivalent to time, and hence only one of them needs to be considered in addition to the two transverse coordinates). 

Even with the above approximation, there are in general no exact solutions for the envelope of the beam. However, we might not even be interested in this exact solution, but instead in its dependence on a certain set of macroscopic parameters which have a clear physical interpretation or can be easily analyzed and controlled in an experiment. Examples of such parameters are the spot size of the beam, its centroid, or its phase. To this end, several methods have been developed and widely applied to the study of Gaussian beams, namely, the source-dependent expansion method~\cite{sprangle1987prl,sprangle1987pra,esarey1994,esarey1997}, the moment method~\cite{lam1975,lam1975}, fully numerical methods~\cite{Sun1987}, and the variational method~\cite{Anderson1979,duda1999,Duda2000,RenDudaHemker2001,Ren2000,RenDuda2001,Ren2002}. We shall employ the last in this Paper.  It has been extensively used for Gaussian beams, in media both with and without noninstantaneous nonlinearities, leading to self-focusing, self-phase modulation, spot size self-modulation, and centroid hosing of a single beam~\cite{Anderson1979,duda1999,Duda2000,RenDudaHemker2001}, as well as the interaction of two beams~\cite{Ren2000,RenDuda2001,Ren2002,Dong2002,Wu2004}, which leads to mutual attraction and spiralling, braiding, and merging. The interaction of any number of Gaussian beams in simple configurations has also been considered~\cite{Ren2002}.

The Gaussian beam is only the fundamental mode in the expansion of an arbitrary beam, and explicit expressions for the critical power for self-focusing of arbitrary higher-order transverse modes, free or interacting, are also of interest. In particular, we consider the decomposition of an arbitrary beam with circular symmetry into Laguerre-Gauss (LG) modes, which are especially relevant since they carry OAM. LG modes are characterized by two integers (see Section~\ref{sec:singleLG} below): the radial index $p\geq 0$ and the azimuthal index $\ell$, which is related to the vortex structure of the beam and directly quantifies its OAM. 

The special case of $p=0$ consists of a radial profile with a single ring, and its critical power has been addressed in Refs.~\cite{kruglov1992,vuong2006}. Assuming the beam maintains an LG profile with OAM $\ell$ throughout the focusing process (what is called a self-similar collapse or aberrationless approximation) Kruglov et al.\ in Ref.~\cite{kruglov1992} directly integrated the NLSE and extracted the critical power from the conditions of existence of periodic solutions. Such a procedure is not easily generalized to profiles with higher $p$ (i.e.\ more radial nodes). By employing the variational method, Chen and Wang were able to obtain the critical power for propagation in a cubic-quintic medium~\cite{chen2010}. Although not explicitly written out, the critical power for an LG mode with arbitrary $p$ propagating in a Kerr medium can be obtained from the results of Ref.~\cite{chen2010}. The assumption of self-similar collapse is also built in the variational method (for a discussion see, e.g., Ref.~\cite{desaix1991}) and hence used throughout this Paper. In Refs.~\cite{fibich2000,fibich2008PRA,fibich2008,fibich2015}, it was shown that when the collapse is not self-similar the analytic prediction for the critical power given in Ref.~\cite{kruglov1992} becomes an upper bound for the critical power. For non-self-similar collapse, only for certain input vortex profiles can an analytic estimate be given for the critical power. Unfortunately, this is not the case for the LG modes. 

Besides the analytical works already mentioned, the propagation and stability of LG modes and other types of vortex beams (e.g., Airy vortex beams~\cite{chen2010airy,chen2013,chen2015,chen2016}) have also been extensively studied numerically, in instantaneous Kerr~\cite{kruglov1985,kruglov1988,soljacic1998,soljacic2000,soljacic2001,grow2007,ishaaya2008}, nonlocal~\cite{buccoliero2007,zhong2007,shen2011}, and saturable~\cite{firth1997,skryabin1998,desyatnikov2001,bigelow2002,desyatnikov2002,bouchard2016} media; see also Ref.~\cite{desyatnikov2004} for a review. In particular, Refs.~\cite{desyatnikov2001,bigelow2002,bouchard2016} considered incoherent superpositions of two vortices.

In this Paper, we use the variational method to analytically study an arbitrary superposition of LG modes of any order, neglecting interference effects between them (this condition can be met, for example, when the phase of each mode varies arbitrarily). This analysis leads to a rich phenomenology of focusing phenomena, with the following two main novel findings. First, we show that an equilibrium configuration where all modes evolve with stationary spot size may not always exist, a fundamental difference regarding single-beam propagation where matched spot size evolution can always be attained. Second, we show that in some cases the total power required for self-focusing is lower for a set of interacting beams than for those beams propagating alone, which is the case of a high OAM mode being guided by low OAM modes. Since it is possible to fully sort LG modes~\cite{fu2018}, the results of this work should be important for the guided propagation of intense pulses, with particular implications in compact laser-plasma accelerators and optical communications.

The Paper is organized as follows. In Section~\ref{sec:review_variational} we briefly review the variational method. As a steppingstone to interacting LG modes, we apply it to obtain the explicit formula for the self-focusing critical power of a single LG beam with arbitrary $p$ (Section~\ref{sec:singleLG}), relating it to known results in the literature. Section~\ref{sec:interactingLG} contains our central result, the critical power for an incoherent superposition of LG beams, Eq.~(\ref{eq:critical_power_interactions}), and a discussion of the rich associated phenomenology. We draw our conclusions and discuss possible applications and extensions to this work in Section~\ref{sec:conclusions}. For completeness, and as an illustration of the power of the variational method, in Appendix~\ref{sec:HGbeams} we derive the critical power for arbitrary Hermite-Gauss modes (in a superposition or not), for which no known expressions exist in the literature, to the best of our knowledge. Appendices~\ref{app:Algebraic_steps_LG}~and~\ref{app:Integral_properties} present algebraic details and useful properties of special functions, respectively.

\section{Brief review of the variational method}\label{sec:review_variational}
Considering the instantaneous response of an optical medium with Kerr nonlinearity, the paraxial evolution of the envelope of a laser is described by the nonlinear Schr{\"o}dinger equation
\begin{equation}\label{eq:NLS}
    \left(2ik_0 \frac{\partial}{\partial z}+\nabla_\perp^2+2\kappa^2\abs{a}^2\right)a=0\,,
\end{equation}
where $a$ is the linearly polarized, normalized envelope of the vector potential~$A$,
\[\frac{1}{2}a(\vec{r}_\perp,z,t)\exp{-ik_0(ct-z)}+\mathrm{c.c.}\,=\frac{e}{mc^2}A(\vec{r}_\perp,z,t)\,,\]
$e$, $m$, $c$ are the electron charge, electron mass, and speed of light, respectively, $k_0$ is the laser wave number, $\kappa^2$ is a medium-dependent constant proportional to the nonlinear part of the refractive index (e.g., $\kappa^2\equiv k_p^2/8$ in a plasma, with $k_p^2$ the plasma wave number), $z$ is the coordinate along the beam, $\vec{r}_\perp$ are the coordinates transverse to $z$, and $\nabla^2_\perp$ is the Laplacian in the transverse plane. The paraxial wave equation is derived in speed of light frame variables $\tau\equiv z$, $\psi\equiv ct-z$. Yet, a single longitudinal variable $z$ suffices here, since we are assuming translational invariance along the beam. Some media, such as plasmas, have a minimum propagation (cutoff) frequency $\omega_p<\omega_0=k_0 c$ (in the plasma, $\omega_p$ is the electron plasma frequency), and in such cases an extra term $-k_p^2 a$ exists inside the bracket of Eq.~(\ref{eq:NLS}). In this type of media, the envelope description (and hence this work) is valid only if the frequency of the beam is considerably larger than the cutoff frequency. However, the extra term in the NLSE just contributes with an overall constant in our Lagrangians below, and hence we simply drop it throughout.

Equation (\ref{eq:NLS}) can be obtained by minimizing the action $S=\int \mathcal{L}\,dz d\vec{r}_\perp$, where the appropriate Lagrangian density is
\begin{equation}\label{eq:Lagrangiandensity}
    \mathcal{L}=ik_0\left(a \frac{\partial a^*}{\partial z}-a^*\frac{\partial a}{\partial z}\right)+\vec{\nabla}_\perp a^*\cdot\vec{\nabla}_\perp a-\kappa^2 a^2{a^*}^2,
\end{equation}
using the Euler-Lagrange equations
\begin{equation*}
    \frac{\partial}{\partial z}\left(\frac{\partial\mathcal{L}}{\partial(\partial a^*/\partial z)}\right)+\vec{\nabla}_\perp\cdot\left(\frac{\partial\mathcal{L}}{\partial(\vec{\nabla}_\perp a^*)}\right)-\frac{\partial \mathcal{L}}{\partial a^*}=0\,.
\end{equation*}
Instead of using the action in terms of the Lagrangian density $\mathcal{L}$ to solve the problem exactly, which is in general not possible, we seek an approximate solution. To that end, we can make an ansatz for the functional form of the envelope, introducing a set of parameters $\beta_i(z)$ depending only on $z$ which fully characterize the envelope $a=a(\beta_i)$ and define a reduced Lagrangian $L(\beta_i)=\int \mathcal{L}\,d\vec{r}_\perp$. The set of parameters act as generalized coordinates, with respect to which the action $S=\int L\,dz$ can be varied. The resulting reduced Euler-Lagrange equations,
\begin{equation*}
    \frac{d}{dz}\left(\frac{\partial L}{\partial \dot{\beta_i}}\right)-\frac{\partial L}{\partial \beta_i}=0\,,
\end{equation*}
(with $\dot{\beta_i}=d\beta_i/dz)$, determine the evolution of the parameters, thus fully characterizing the evolution of the envelope~$a$. By integrating out the transverse coordinates from the action we are thus substituting an infinite number of (transverse) degrees of freedom in the field~$a$ by a finite number of mechanical coordinates~$\beta_i$, reducing the problem to a system of coupled ordinary differential equations. In principle, the solution can be made arbitrarily exact by considering more parameters $\beta_i$, at the cost of computational complexity. The degree of approximation of the variational method depends, therefore, on the choice of trial function.

\section{Self-focusing of a single Laguerre-Gauss mode}\label{sec:singleLG}
We start by considering the $(p,\ell)$ Laguerre-Gauss mode as our trial function,
\begin{equation}\label{eq:trialfunction}
\begin{split}
    a=&\,A\,C_{p\ell}\,\left(\frac{\sqrt{2}r}{W}\right)^{\abs{\ell}}L_p^{\abs{\ell}}\!\left(\frac{2r^2}{W^2}\right)\exp{-\frac{r^2}{W^2}}\times\\
    \times&\,\exp{i\left(\ell\varphi+k_0\frac{r^2}{2R}-\psi\right)}\,,
\end{split}
\end{equation}
where the parameters~$\beta_i$ are the amplitude~$A$, the spot size~$W$, the radius of curvature~$R$ and the phase~$\psi$. The transverse coordinates are polar $(r,\varphi)$, with the beam centered at the origin, $p$ is the radial mode number (it gives the number of radial nodes), $\ell$ is the azimuthal mode number or OAM of the vortex, and $L_p^{\abs{\ell}}$ is an associated Laguerre polynomial. The normalization constant $C_{p\ell}=\sqrt{p!/(p+\abs{\ell})!}$ is such that the power~$P$ of each mode is $P=(2/\pi)\int\abs{a}^2d\vec{r}_\perp=A^2W^2$ as is usually done for Gaussian beams. Other parameters could be considered, e.g., the centroid of the beam and some momenta transverse to the propagation axis (i.e.\ components $\vec{k}_\perp\perp\vec{k}_0$), but they are not crucial in what follows, since we consider only different beams centered at the origin as is appropriate for the modes resulting from a decomposition of an arbitrary beam. However, these parameters have to be considered if one wishes to study the interaction of several beams at different transverse locations.

Inserting the trial function into Eq.~(\ref{eq:Lagrangiandensity}) and integrating over all $r$ and $\varphi$ we obtain the reduced Lagrangian (additional details are presented in Appendix~\ref{app:Algebraic_steps_LG})
\begin{equation}\label{eq:Lagrangian}
\begin{split}
    L=&\frac{\pi}{2}A^2C_{p\ell}^2W^2\Biggl[-2I_{100}k_0\dot{\psi}+\frac{1}{2}I_{101}k_0^2W^2\left(\frac{1}{R^2}-\frac{\dot{R}}{R^2}\right)+\\
    +&\frac{4\ell^2}{W^2}I_{10-1}-\frac{4\abs{\ell}}{W^2}(I_{100}+2I_{110})+\\
    +&\frac{8}{W^2}\left(\frac{1}{4}I_{101}+I_{111}+I_{121}\right)-\kappa^2I_{200}A^2C_{p\ell}^2\Biggr],
\end{split}
\end{equation}
where $I_{mns}$ is the following integral, which depends only on the beam mode:
\begin{equation}\label{eq:Idefinition}
    I_{mns}(p,\ell)=\int_0^\infty e^{-mx}x^{m\abs{\ell}+s}\left[L_p^{\abs{\ell}}(x)\right]^{2m-n}\left[L_{p-1}^{\abs{\ell}+1}(x)\right]^{n}dx\,.
\end{equation}
The evaluation of the relevant integrals (using the properties in Appendix~\ref{app:Integral_properties}) yields
\begin{equation}\label{eq:IIntegrals}
\begin{split}
    &I_{100}=\frac{(p+\abs{\ell})!}{p!}\,,\quad I_{101}=\frac{(p+\abs{\ell})!}{p!}(2p+\abs{\ell}+1)\,,\\
    &I_{10-1}=\frac{1}{\abs{\ell}}\frac{(p+\abs{\ell})!}{p!}\,,\quad I_{110}=0\,,\\ 
    &I_{121}=-I_{111}=\frac{(p+\abs{\ell})!}{(p-1)!}\,,\\
    &I_{200}=\frac{1}{2^{4p+2\abs{\ell}+1}}\left[\frac{(p+\abs{\ell})!}{p!}\right]^2S_{p\ell}\,,\\
    &S_{p\ell}=\sum_{n=0}^p\frac{(2n)!\left[(2p-2n)!\right]^2(2\abs{\ell}+2n)!}{(n!)^2\left[(p-n)!\right]^4\left[(\abs{\ell}+n)!\right]^2}\,.
\end{split}
\end{equation}

We now apply the Euler-Lagrange equations to the Lagrangian of Eq.~(\ref{eq:Lagrangian}). Variation with respect to the phase $\psi$ gives rise to power conservation, since
\begin{equation}
    \frac{d}{dz}\left(\frac{\pi}{2}W^2A^2C_{p\ell}^2I_{100}\right)=0\implies\frac{dP}{dz}=0\,.
\end{equation}
Since the variable $A$ arises in the Lagrangian only through the combination $A^2W^2$, we can replace it by the constant $P$, therefore avoiding variations with respect to $A$. Variation with respect to the radius of curvature $R$ relates $R$ to $W$ through $R=W/\dot{W}$. Both this auxiliary condition and power conservation are also present in the Gaussian case. Using these intermediate results, as well as Eqs.~(\ref{eq:IIntegrals}), the variation with respect to the spot size gives the equation for spot size dynamics:
\begin{equation}\label{eq:variationW}
    \ddot{W}+\frac{4}{k_0^2W^3}\left[\frac{\kappa^2P}{4}\frac{2I_{200}C_{p\ell}^2}{I_{101}}-1\right]=0\,.
\end{equation}
A stationary spot size is obtained when the term inside brackets vanishes, i.e.\ for $P=P_c$ where $P_c$ is the critical power given by
\begin{equation}\label{eq:criticalpower}
\begin{split}
    P_c&=\frac{4}{\kappa^2}\frac{I_{101}}{2I_{200}C_{p\ell}^2}=P_G\frac{1}{2I_{200}}\!\left[\frac{(p+\abs{\ell})!}{p!}\right]^2\!(2p+\abs{\ell}+1)\\
    &=P_G\,4^{2p+\abs{\ell}}(2p+\abs{\ell}+1)S_{p\ell}^{-1}\,,
\end{split}
\end{equation}
where $P_G=4/\kappa^2$ is the critical power for a Gaussian beam, which is corrected by a factor characteristic of each LG mode. By setting $p=\ell=0$ in Eq.~(\ref{eq:criticalpower}) we recover $P_c=P_G$ as we should. For the particular case where the laser intensity profile consists of a single ring ($p=0$), we find $I_{200}=4^{-\abs{\ell}}(2\abs{\ell})!/2$ and consequently recover the known result~\cite{kruglov1992}:
\begin{equation}
   P_c=P_G\,4^{\abs{\ell}}\frac{\abs{\ell}!(\abs{\ell}+1)!}{(2\abs{\ell})!}\,.
\end{equation}
The expression for arbitrary $p$, Eq.~(\ref{eq:criticalpower}), agrees with the result of Ref.~\cite{chen2010}, once model-specific constants are related and the quintic medium constant is set to zero.

\begin{figure}[tbp]
  \centering
  \includegraphics[width=0.45\textwidth]{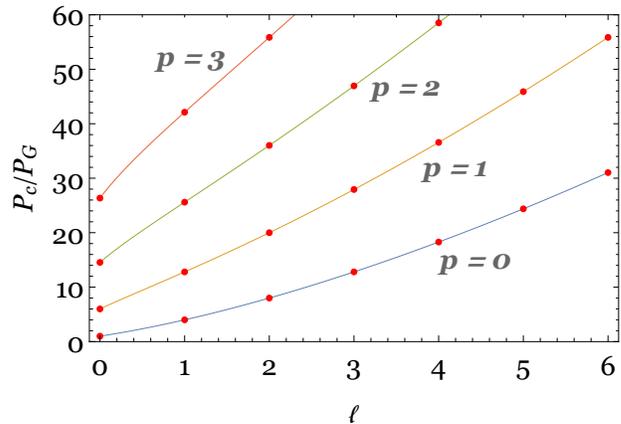}
  \caption{Critical power for different values of radial number for the LG modes. The red dots correspond to integer values of $\ell$. Non-integer values of $\ell$ can be considered, using the $\Gamma$ function, to facilitate visualization. Increasing OAM or radial numbers raises the power threshold.}
  \label{fig:Pc_LG}
\end{figure}

The evolution of the critical power with OAM for various values of $p$ is presented in Fig.~\ref{fig:Pc_LG}. It shows that the critical power rises monotonically with the OAM of the beams. The critical power grows with the radial number $p$. As a result, the mode which is most easily focused is the fundamental Gaussian mode. The fact that the critical power rises with $\ell$ can be understood heuristically using a simple physical picture, based on the centrifugal force felt by the photons of the OAM beam. The OAM beams have helical wavefronts, whereby photons at radius $r$ undergo azimuthal motion in the transverse plane with projected velocity $v/c=k_\perp/k_0$ where $k_\perp\sim-i\vec{\nabla}_\perp\sim\ell/r$ is transverse momentum. In their frame of reference, the photons are thus subjected to an outward centrifugal force $\abs{F}\sim k_\perp^2/r$. For a ring-shaped beam ($p=0$) the profile is peaked at $(r/W)^2=\abs{\ell}/2$ and we have $F\sim\sqrt{\abs{\ell}}$. Hence, the higher the OAM mode, the more the photons are pushed out, and the more difficult it is to focus them inwards, requiring a higher focusing power.

For systems with rectangular instead of circular symmetry, the expansion of a beam is best done in terms of Hermite-Gauss (HG) modes. Proceeding in the same way, one obtains the critical power for HG beams; see Appendix~\ref{sec:HGbeams}.

\section{Self-focusing of incoherent interacting Laguerre-Gauss modes}\label{sec:interactingLG}
A general treatment of interacting higher-order modes must include the interference between them and is not readily treated by the variational method. We will consider only the incoherent case, where we neglect all interference between beams. This approximation is reasonable if the beams have (slightly) different frequencies or, for a sufficient number of different modes, random initial phases, whereby in both cases the interference would be averaged out during the propagation. If we consider only two beams, the incoherent approximation becomes exact if the beams have orthogonal polarizations. We decompose the beam envelope $a=\sum_i a_i$ into a linear combination of LG modes $a_i$, given by Eq.~(\ref{eq:trialfunction}) with an index $i$ in the parameters $A$, $W$, $R$, and $\psi$ and mode numbers $(p_i,\ell_i)$. The amplitudes $A_i$ of each mode act as linear coefficients in the expansion. Then, the incoherent approximation leads to $\abs{a}^2\approx\sum_i \abs{a_i}^2$, which can be substituted in Eq.~(\ref{eq:NLS}) for each mode envelope $a_i$. Hence, the interaction of incoherent beams is accounted by an extra term $2\kappa^2 \sum_{j\neq i}\abs{a_j}^2a_i$ in the NLSE or equivalently by an extra pairwise amplitude-amplitude interaction term in the Lagrangian density~\cite{Ren2002},
\begin{equation}\label{eq:interaction_Lagrangian}
    \mathcal{L}_\mathrm{int}=-\kappa^2\sum_{i,j\neq i}a_i^*a_ia_j^*a_j=-\kappa^2\sum_{i,j\neq i}\abs{a_i}^2\abs{a_j}^2\,.
\end{equation}

We can again understand the form of the interaction in terms of a physical picture. A relativistic beam will increase locally the index of refraction~$n$ of the medium according to $n=n_0+\Delta n$, with $n_0$ the linear index of refraction and $\Delta n\propto\kappa^2\abs{a}^2$. LG modes propagate in the $z$-direction, have a gradient of $n$ in the $r$ direction, and hence are attracted towards radii of higher $n$. When two beams interact, the peaks of the first attract the second and vice versa, while the nodes of intensity do not attract at all, in accordance with Eq.~(\ref{eq:interaction_Lagrangian}). In the case of an OAM beam, this effective attractive force has to compete with the centrifugal force discussed previously. This physical picture proves valuable in analyzing the results at the end of this section.

Since the new interaction term does not depend on the phases of the modes, only the equation for the spot sizes is modified. Setting $W_i=W$ for all modes after differentiating, the variation with respect to $W_i$ results in an extra term
\begin{equation}\label{eq:interaction_correctionn_EoM}
\begin{split}
    \frac{\partial L_\mathrm{int}}{\partial W_i}&\bigg\rvert_{W_i=W}=\frac{4\kappa^2}{W_i^3}\left(\frac{\pi}{2}P_iC_{p_i\ell_i}^2\right)\\
    &\times\sum_{j\neq i}P_jC_{p_j\ell_j}^2(\abs{\ell_i}H^{ij}_{00}-H^{ij}_{01}-2H^{ij}_{11})\,,
\end{split}
\end{equation}
with the $H_{ns}$ integrals defined analogously to before:
\begin{equation}
\begin{split}
    H^{ij}_{ns}(p_i,\ell_i,p_j,\ell_j)=\int_0^\infty e^{-2x}x^{\abs{\ell_i}+\abs{\ell_j}+s}\left[L_{p_i}^{\abs{\ell_i}}(x)\right]^{2-n}\\
    \times\left[L_{p_i-1}^{\abs{\ell_i}+1}(x)\right]^{n}\left[L_{p_j}^{\abs{\ell_j}}(x)\right]^{2}dx\,.
\end{split}
\end{equation}

Adding Eq.~(\ref{eq:interaction_correctionn_EoM}) to Eq.~(\ref{eq:variationW}), leads to a corrected equation of motion for the spot size of each mode given by
\begin{equation}\label{eq:eom_W_interactions}
\begin{split}
    \ddot{W_i}&+\frac{4}{k_0^2W_i^3}\Bigg[\frac{\kappa^2P_i}{4}\frac{2I^i_{200}C_{p_i\ell_i}^2}{I^i_{101}}\\
    &+\kappa^2\sum_{j\neq i}P_jC_{p_j\ell_j}^2\frac{\abs{\ell_i}H^{ij}_{00}-H^{ij}_{01}-2H^{ij}_{11}}{I_{101}^i}-1\Bigg]=0\,.
\end{split}
\end{equation}
Equating the term inside square brackets to zero yields the critical power for each beam, while the set of powers which leads to all modes evolving with constant $W_i$ is the solution to the following system of linear equations:
\begin{equation}\label{eq:critical_power_interactions}
    \sum_j\!\left[\delta_{ij}+(1-\delta_{ij})\,2\,\frac{C_{p_j\ell_j}^2}{C_{p_i\ell_i}^2}\frac{\abs{\ell_i}H^{ij}_{00}-H^{ij}_{01}-2H^{ij}_{11}}{I_{200}^i}\right]\!P^c_j=P_i^0\,,
\end{equation}
where $P_i^0$ is the free critical power given by Eq.~(\ref{eq:criticalpower}) and $\delta_{ij}$ is the Kronecker delta.

Equation (\ref{eq:critical_power_interactions}) admits solutions where the waists of all beams propagate with constant spot size $W_i$. This kind of solution represents an equilibrium configuration where all spot sizes are stationary. In addition to these self-focusing solutions, Eq.~(\ref{eq:critical_power_interactions}) also predicts the existence of mode combinations where no stationary solution can be found. This is in stark contrast with the single-mode analysis that characterizes the usual self-focusing theory, valid for a single beam.

We start with the cases where there is no equilibrium. For some combinations of interacting modes, a number of $P^c_j$ come out negative, and since power is non-negative one concludes that no physical solutions exist and hence the equilibrium cannot be attained. In terms of the above physical picture, one interprets this as an impossibility to exactly balance all the attractions of peaks and the centrifugal forces on the beams, and to attain the equilibrium one would require repulsive forces (expressed as negative power), to stabilize these configurations. Examples of this situation are the interaction of three beams with $p=0$ and $\ell=0$, $\pm1$ with the $(1,0)$-mode, in which all four required powers are negative, and the interaction of five beams with $p=0$ and $\ell=0$, $\pm1$, $\pm2$ in which only the $\ell=\pm2$ would require negative power. Let us be more specific. For any of these cases, if one inserts any positive (that is, physical) set of values for powers $P_i$ into Eq.~(\ref{eq:eom_W_interactions}), then one obtains at least one $W_i>0$, i.e.\ at least one beam defocuses. By varying $P_i$ smoothly, all $\ddot W_i$ will also vary smoothly. But we have seen that one can never go through the equilibrium configuration $\ddot W_i=0$ for all $i$ (because the powers required for that laid outside the set of physical solutions). So, we cannot reach $\ddot W_i<0$ (all beams focus), because any smooth variation would have to go through $\ddot W_i=0$. We thus conclude that for a set of interacting higher-order modes, it may not be possible to reach self-focusing propagation simultaneously for all beams. This result is fundamentally different from the single-beam case, where self-focusing is always possible, given high enough power.

In the cases where the equilibrium configuration does indeed exist, it is not enough to raise slightly the power of one beam to focus it, since each power now depends on the values of all others, and changing one could break the equilibrium of the others. One has to define a perturbation of the power for all beams and reinsert it into Eq.~(\ref{eq:critical_power_interactions}). We denote the term inside square brackets in Eq.~(\ref{eq:critical_power_interactions}) by $M_{ij}$ and perturb $P_j=P^c_j+\delta\!P_j$. Then the $i$-th beam will focus if $\sum_jM_{ij}\,\delta\!P_j$ is positive and will defocus if it is negative (the limiting case $\sum_jM_{ij}\,\delta\!P_j=0$ corresponds to the equilibrium configuration). To see this, we note that, using Eq.~(\ref{eq:critical_power_interactions}) written as $\sum_{ij}M_{ij}(P_j^c+\delta\!P_j)=P^0_j$, Eq.~(\ref{eq:eom_W_interactions}) can be rewritten as
\begin{equation}
    \ddot{W_i}+\frac{4}{k_0^2W_i^3}\frac{\kappa^2}{4}\frac{2I^i_{200}C_{p_i\ell_i}^2}{I^i_{101}}\sum_jM_{ij}\,\delta\!P_j=0\,.
\end{equation}
Since the prefactor of $\sum_jM_{ij}\,\delta\!P_j$ is always positive, focusing (resp.\ defocusing) of the $i$-th beam, i.e.\ $\ddot{W}_i<0$ (resp.\ $\ddot{W}_i>0$), occurs if $\sum_jM_{ij}\,\delta\!P_j>0$ (resp.\ $<0$). If we can find a vector $\delta\!P$ such that all of the entries of $M\cdot\delta\!P$ are positive, then it is possible to focus all of the beams simultaneously. This procedure is illustrated in several examples to follow.

\begin{figure}[tbp]
    \centering
    \includegraphics[width=0.45\textwidth]{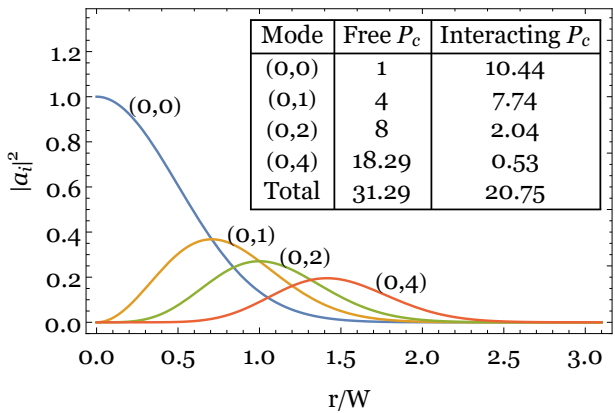}
    \caption{Radial profile for the interacting $(p,\ell)$ modes with $p=0$ and $\ell=0$, $1$, $2$, $4$ and corrections to the critical powers for these modes due to the interaction. Both the free critical powers, given by Eq.~(\ref{eq:criticalpower}), and the interacting critical powers, given by the solutions to Eq.~(\ref{eq:critical_power_interactions}), are given in multiples of $P_G$. Due to the outward net force, the inner beams, with $\ell=0$, $1$ become harder to focus ($P_c$ increases) while the beams with $\ell=2$, $4$ are attracted inwards and $P_c$ becomes smaller. In particular for the $\ell=4$ beam the interacting $P_c$ is smaller than 1, which is the free $P_c$ for a Gaussian beam. Since the decreases in $P_c$ for the $\ell=0$, $2$ modes are larger than the increases for the $\ell=0$, $1$ beams, the total power required for stationary evolution goes down from $31.29P_G$ to $20.75P_G$.}
    \label{fig:interaction_Pc}
\end{figure}

A remarkable result is that the sum of the critical powers for all the beams can be less when they interact than when they do not. Also, for free modes it was seen that the Gaussian beam always has the lowest critical power. However, in the presence of interactions, this no longer holds. An example illustrative of both these results is the interaction of four modes with $p=0$ and $\ell=0$, $1$, $2$, $4$, whose profiles are depicted in Fig.~\ref{fig:interaction_Pc}, so that the physical picture can be visualized, together with their free and interacting critical powers. In this case, any perturbation around the equilibrium in which the powers of the $\ell=0$, $1$ beams are slightly higher than the critical power will lead to focusing of all four beams. Thus, it is possible to focus the interacting beams with only around two-thirds of the power required for focusing the four beams if they were propagating individually through the medium. Note also that, in this superposition, the Gaussian beam no longer has the lowest critical power, having indeed the highest.

Even more interesting is the possibility of using low OAM modes to help focus a high OAM beam. For instance, the mode with $p=0$ and $\ell=10$ has a (free) critical power $P_c\approx 62.43P_G$. However, if we consider the simultaneous propagation of this mode together with the three modes with $p=0$ and $\ell=0,5,8$, the total power required to focus \emph{all four} beams at once is only $P_c\approx 37.87P_G$. Hence, it is energetically favorable to focus this set of beams instead of the $\ell=10$ beam individually, i.e.\ there is power saved by injecting further modes in the medium. In principle, by choosing modes with higher $p$~and~$\ell$ and combinations with a higher number of modes, the decrease in critical power can be made arbitrarily large. However, the final required power would still be very high (although much less than the free power) since the critical power rises very fast with increasing mode. Nonetheless, it could be that, in some cases, the interactions bring the critical power from above the current technological capabilities to below them; the above, complicated, combinations may, then, prove to be important.

\begin{figure}[tbp]
    \centering
    \includegraphics[width=0.45\textwidth]{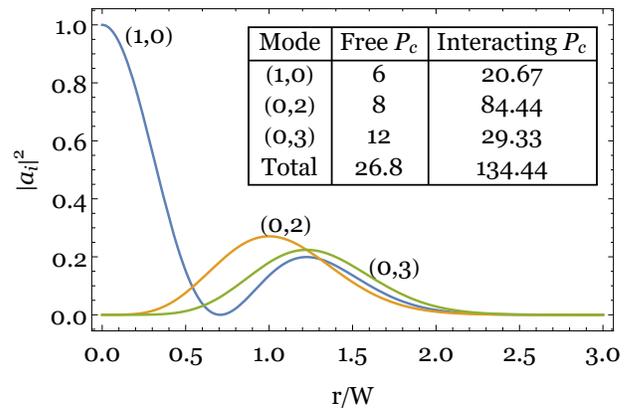}
    \caption{Radial profile for the interacting $(1,0)$, $(0,2)$ and $(0,3)$ modes and corrections to the critical powers for these modes due to the interaction. Free and interacting powers are again multiples of $P_G$. Effective forces are such that all beams spread: the innermost $(1,0)$ mode is pulled by the external peaks of the $p=0$ modes, while those modes feel an effective inward force due to the first peak of the $(1,0)$ mode and an effective outward force due to the second peak of the $(1,0)$ mode and the other $p=0$ mode. The total power required for stationary evolution goes up from $26.8P_G$ to $134.44P_G$.}
     \label{fig:interaction_Pc_2}
\end{figure}

In some other cases, all of the critical powers increase. As an example, the interaction of the $(1, 0)$, $(0, 2)$, and $(0, 3)$ modes, has a fivefold increase in critical power required as is depicted in Fig.~\ref{fig:interaction_Pc_2}. It is again qualitatively explained by the physical picture. As before, the increase of critical power can be made arbitrarily large. These cases are also of considerable importance when one wants the beams not to focus, and the calculations of critical power must include the interactions to ensure that the power is in fact below the threshold. 

Given an arbitrary beam envelope with circular symmetry, one can decompose it into LG modes, and knowing the spot size of the beam, the coefficients of the expansion squared give the power of each mode, which can then be compared to the solutions of Eq.~(\ref{eq:critical_power_interactions}) to check if the beam as a whole will focus or not, as long as the modes can be considered incoherent superpositions. This analysis results from a situation where one solves Eq.~(\ref{eq:critical_power_interactions}) for all beams involved in the interaction. However, it is also possible to tune and fix the power of one (or more) of the modes to increase or decrease $P_c$ for the remaining ones. In this case, only a subsystem of Eq.~(\ref{eq:critical_power_interactions}) is to be solved.

\section{Conclusions}\label{sec:conclusions}
In summary, we have obtained explicit expressions for the critical self-focusing power of all basis modes with circular symmetry in a Kerr medium. Here, the main conclusion is that the fundamental Gaussian mode has the lowest critical power, which increases monotonically with both $p$~and~$\ell$. In addition, we have shown that, as was already known for Gaussian beams~\cite{Ren2000,RenDuda2001,Ren2002}, LG modes can attract each other, altering the power thresholds for simultaneously propagating beams and for different modes of a decomposition of an arbitrary beam. Although the interaction is always attractive, the existence of multiple peaks and troughs in the intensity profile of LG beams allows for combinations of modes where the interaction leads to focusing or defocusing of some or all beams. Together with the natural diffraction of the laser beams and the enhanced defocusing of LG modes due to its OAM, this allows for rich possibilities of focusing phenomena. In particular, in major contrast with single-beam propagation, propagation for all beams with stationary spot size may be impossible to attain.

Our results could prove useful in a variety of situations. First, if we want to perform experiments with multiple modes where self-focusing is important, one can perform the experiments simultaneously, saving power in the focusing process,  since it is now possible to fully sort LG modes~\cite{fu2018}. Second, even if one is interested in the propagation of a single beam (in particular, a high OAM one), it may be energetically favorable to use other beams (particularly low OAM ones) to help focus it. Third, when one wants the beams \emph{not} to focus (for instance, to prevent optical damage to the nonlinear medium), secondary beams could be used to raise the critical power of the main beam, avoiding self-focusing.

The quantitative expressions derived in this work should be most reliable at the onset of self-focusing and for moderate powers (not too high compared with $P_c$). On the one hand, this ensures that higher-order corrections in the NLSE may be neglected and also that the paraxial approximation holds. On the other hand, it has been pointed out~\cite{chen2010} that the results of the variational method are most accurate for moderate powers. This shortcoming at intense powers could, in principle, be overcome by choosing a trial function with more variational parameters, and therefore does not rule out the variational method itself as an accurate approach to self-focusing. Notwithstanding the (controllable) loss of quantitative accuracy in certain situations, the qualitative picture introduced in this work remains valuable. 

Furthermore, the set of variational parameters considered in this Paper allows for the study of self-focusing only. The inclusion of centroid positions and transverse momenta would allow other possibilities, namely, hosing of a single beam or interaction of several beams at different transverse positions, as has been done for the Gaussian case. Furthermore, this work considered only instantaneous quadratic nonlinearities, neglecting spatiotemporal evolution along the beam (e.g., the coupling of a laser to plasma waves). The Lagrangian density for this more general case is known~\cite{Duda2000,RenDuda2001,Ren2002}, but a study of this phenomenon for higher-order modes is still lacking. Also of interest would be to extend the variational method to include interference between modes.  In that case, power would no longer be constant and could be exchanged between the beams, and dynamical equations for the amplitude and the phase would need to be taken into account. The phenomenon of filamentation could possibly be treated this way.

Finally, other trial functions (say, Hermite-Gauss, Bessel, or Airy beams) could be employed to $(i)$ obtain the respective critical powers and $(ii)$ investigate some new possible phenomenology arising due to interactions. We followed path $(i)$ for Hermite-Gauss beams, and the expressions thus obtained are given in Appendix~\ref{sec:HGbeams}.

\acknowledgments
We acknowledge fruitful discussions with Luis Oliveira e Silva. We thank the anonymous reviewer whose comments have greatly helped improve this work. We acknowledge the EU Accelerator Research and Innovation for European Science and Society (EU ARIES) Grant Agreement No.~730871 (H2020-INFRAIA-2016-1). J.~V.\ acknowledges the support of FCT (Portugal) Grant No.~SFRH/IF/01635/2015.

\appendix

\section{Self-Focusing of Hermite-Gauss Beams}\label{sec:HGbeams}
For completeness, we now consider the case where a system exhibits rectangular instead of cylindrical symmetry. The beam expansion is then most conveniently done in terms of Hermite-Gauss (HG) modes. The trial function is the $(m,n)$ HG mode,
\begin{equation}\label{eq:trialfunction_HG}
\begin{split}
    a=&\,A\,C_{mn}\,H_m\left(\frac{\sqrt{2}x}{W}\right)H_n\left(\frac{\sqrt{2}y}{W}\right)\exp{-\frac{x^2+y^2}{W^2}}\times\\
    \times&\,\exp{i\left(k_0\frac{x^2+y^2}{2R}-\psi\right)}\,,
\end{split}
\end{equation}
where $A$, $W$, $R$, and $\psi$ are as above, the transverse coordinates are Cartesian $(x,y)$, $m$ and $n$ are the $x$ and $y$ mode numbers, respectively, and give the number of nodes in each direction, and $H_n$ is a Hermite polynomial. The normalization constant $C_{mn}=1/\sqrt{2^{m+n}m!n!}$ is again chosen such that $P=A^2W^2$, irrespective of the mode.

Following the same procedure as in Section~\ref{sec:singleLG}, the integrated Lagrangian is found to be 
\begin{equation}\label{eq:Lagrangian_HG}
\begin{split}
    L=&\frac{1}{2}A^2C_{mn}^2W^2\Biggl[-2J^m_{100}J^n_{100}k_0\dot{\psi}\\
    +&\frac{1}{2}(J^m_{102}J^n_{100}+J^m_{100}J^n_{102})k_0^2W^2\left(\frac{1}{R^2}-\frac{\dot{R}}{R^2}\right)+\\
    +&\frac{8}{W^2}(m^2J^m_{120}J^n_{100}+n^2J^m_{100}J^n_{120}-mJ^m_{111}J^n_{100}-nJ^m_{100}J^n_{111})\\
    +&\frac{2}{W^2}(J^m_{102}J^n_{100}+J^m_{100}J^n_{102})+-\kappa^2 J^m_{200}J^n_{200}A^2C_{mn}^2\Biggr],
\end{split}
\end{equation}
where the relevant integrals for the HG beams are defined by
\begin{equation}\label{eq:Jdefinition_HG}
    J_{\alpha\beta\gamma}^q=\int_{-\infty}^{+\infty} e^{-\alpha \xi^2}\xi^{\gamma}\left[H_q(\xi)\right]^{2\alpha-\beta}\left[H_{q-1}(\xi)\right]^{\beta}d\xi\,.
\end{equation}
The $\alpha=1$ integrals can be readily evaluated using the orthogonality and recursion relations for Hermite polynomials of Appendix~\ref{app:Integral_properties}:
\begin{equation}\label{eq:JIntegrals_HG}
\begin{split}
    J^q_{100}=\sqrt{\pi}2^qq!\,,\quad J^q_{102}=\left(q+\frac{1}{2}\right)J^q_{100}\,,\\ J^q_{120}=\frac{1}{2q}J^q_{100}\,,\quad J^q_{111}=\frac{1}{2}J^1_{100}\,.
\end{split}
\end{equation}

Variation of the Lagrangian with respect to $\psi$ and $W$ gives the same intermediate results as in Section~\ref{sec:singleLG} and the variation with respect to the spot size yields the HG critical power
\begin{equation}\label{eq:criticalpower_HG}
    P_c=P_G\frac{\pi}{2J^m_{200}J^n_{200}}(2^{m+n}m!n!)^2(m+n+1)\,.
\end{equation}

As in the cylindrical case, the lowest mode ($m=n=0$) recovers the Gaussian result as expected and for higher modes self-focusing is more difficult since the power threshold rises.

\begin{widetext}
The interaction Lagrangian density is again the one of Eq.~(\ref{eq:interaction_Lagrangian}), which after integration in the transverse plane and variation with respect to $W$ leads to the interacting equation of motion for the spot sizes of each beam,
\begin{equation}
\begin{split}
    \ddot{W}+\frac{4}{k_0W^3}\Biggl[&\frac{\kappa^2P_i}{8}\frac{2C_{n_im_i}^2J_{200}^{m_i}J_{200}^{n_i}}{J_{102}^{m_i}J_{100}^{n_i}+J_{100}^{m_i}J_{102}^{n_i}}+\frac{k_p^2}{8}\sum_{j\neq i}P_jC_{m_jn_j}^2\\
    &\times\frac{2m_iK_{11}^{m_im_j}K_{00}^{n_in_j}+2n_iK_{00}^{m_im_j}K_{11}^{n_in_j}-K_{02}^{m_im_j}K_{00}^{n_in_j}-K_{00}^{m_im_j}K_{02}^{n_in_j}}{J_{102}^{m_i}J_{100}^{n_i}+J_{100}^{m_i}J_{102}^{n_i}}    
    -1\Biggr]=0\,,
\end{split}
\end{equation}
where the integrals $K_{\beta\gamma}^{qr}$ are defined as
\begin{equation}
    K_{\beta\gamma}^{qr}=\int_{-\infty}^{+\infty}e^{-2\xi^2}\xi^\gamma\left[H_q(\xi)\right]^{2-\beta}\left[H_{q-1}(\xi)\right]^{\beta}\left[H_{r}(\xi)\right]^{2}d\xi\,.
\end{equation}
Equating the term in square brackets to zero gives the linear system 
\begin{equation}
    \sum_j\left[\delta_{ij}+(1-\delta_{ij})\,2\,\frac{C_{m_jn_j}^2}{C_{m_in_i}^2}\times\frac{2m_iK_{11}^{m_im_j}K_{00}^{n_in_j}+2n_iK_{00}^{m_im_j}K_{11}^{n_in_j}-K_{02}^{m_im_j}K_{00}^{n_in_j}-K_{00}^{m_im_j}K_{02}^{n_in_j}}{J_{200}^{m_i}J_{200}^{n_i}}\right]P_j=P_i^0\,,
\end{equation}
where $P_i^0$ is given by Eq.~(\ref{eq:criticalpower_HG}),
whose solution is the set of interacting critical powers for interacting HG modes.
\end{widetext}

\section{Algebraic computations leading to the reduced LG Lagrangian}\label{app:Algebraic_steps_LG}
In this appendix we present some auxiliary computations leading to the Lagrangian of Eq.~(\ref{eq:Lagrangian}). We start with the first term in the Lagrangian density of Eq.~(\ref{eq:Lagrangiandensity}), noting that $a\,\partial a^*/\partial z-a^*\partial a/\partial z=2i\,\mathrm{Im}(a\,\partial a^*/\partial z)$. Using the trial function of Eq.~(\ref{eq:trialfunction}), we have \begin{equation}
    \mathrm{Im}\left(a\,\frac{\partial a^*}{\partial z}\right)=\abs{a}^2\left(\dot{\psi}+k_0\frac{r^2}{2R^2}\dot{R}\right)\,.
\end{equation}
For the gradient term, we must use the derivative property of associated Laguerre polynomials $(L_p^{\abs{\ell}})'(x)=-L_{p-1}^{\abs{\ell}+1}(x)$, where the prime denotes differentiation with respect to the whole argument. Using the gradient in polar coordinates, $\vec{\nabla}_\perp=\bm{\hat{r}}\partial/\partial r+\bm{\hat{\varphi}}\,1/r\,\partial/\partial\varphi$, we obtain
\begin{equation}
\begin{split}
    \vec{\nabla}_\perp a^*&\cdot\vec{\nabla}_\perp a=\frac{\abs{a}^2}{r^2}\Biggl[\frac{k_0r^4}{R^2}+\ell^2+\\ &+\left(\abs{\ell}-\frac{2r^2}{W^2}\left(1+2\frac{L_{p-1}^{\abs{\ell}+1}(\frac{2r^2}{W^2})}{L_{p}^{\abs{\ell}}(\frac{2r^2}{W^2})}\right)\right)^2\Biggr]\,.
\end{split}
\end{equation}
The last term is trivial, $a^2a^{*2}=\abs{a}^4$. Concerning the integration of $\mathcal{L}$ in the transverse plane, all integrals in $\varphi$ give $2\pi$, and the radial integrals are proportional to either $\int_0^\infty \abs{a}^2rdr$, $\int_0^\infty \abs{a}^4rdr$, $\int_0^\infty \abs{a}^2r^3dr$, $\int_0^\infty \abs{a}^2/rdr$, or $\int_0^\infty \abs{a}^2\,L_{p-1}^{\abs{\ell}+1}(2r^2/W^2)/L_{p}^{\abs{\ell}}(2r^2/W^2)rdr$, which can be brought to the form $I_{mns}$ of Eq.~(\ref{eq:Idefinition}) using the change of variable $x=2r^2/W^2$. Adding all the above terms yields the Lagrangian of Eq.~(\ref{eq:Lagrangian}).

The derivation of the HG Lagrangian is identical, but uses the derivative of the Hermite polynomial, $(H_n)^\prime(x)=2nH_{n-1}(x)$. In this case, the integrals can be separated in $x$ and $y$, and hence every term in Lagrangian of Eq.~(\ref{eq:Lagrangian_HG}) has two $J$-integrals multiplying it.

\section{Useful properties of Special Functions}\label{app:Integral_properties}
The following orthogonality and recursion relations of the associated Laguerre polynomials~\cite{AbramowitzStegun} were used in evaluating the $I$-integrals:
\begin{subequations}
\begin{equation}
   \int_0^\infty e^{-x}x^{\abs{\ell}}L_p^{\abs{\ell}}(x)L_q^{\abs{\ell}}(x)\,dx=\frac{(p+\abs{\ell})!}{p!}\delta_{pq}\,,
\end{equation}    
\begin{equation}
   \int_0^\infty e^{-x}x^{\abs{\ell}+1}\left[L_p^{\abs{\ell}}(x)\right]^2dx=\frac{(p+\abs{\ell})!}{p!}(2p+\abs{l}+1)\,,
\end{equation}
\begin{equation}
   L_p^{\abs{\ell}}(x)=L_p^{\abs{\ell}+1}(x)-L_{p-1}^{\abs{\ell}+1}(x)\,,
\end{equation}
\begin{equation}
   L_p^{\abs{\ell}+1}(x)=\sum_{k=0}^pL_k^{\abs{\ell}}(x)\,.
\end{equation}
\end{subequations}
Also used were the orthogonality and recursion relations for Hermite polynomials, given by
\begin{subequations}
\begin{equation}
   \int_{-\infty}^{+\infty} e^{-\xi^2}H_m(\xi)H_n(\xi)\,d\xi=\sqrt{\pi}2^nn!\,,
\end{equation}    
\begin{equation}
   H_{n+1}(\xi)=2\xi H_m(\xi)-2nH_{n-1}(\xi)\,.
\end{equation}
\end{subequations}

\end{document}